\documentclass[twocolumn,prb,aps,floatfix,showpacs,superscriptaddress]{revtex4}
\usepackage{graphicx}
\usepackage{dcolumn}
\usepackage{bm}

\begin{document}

\title{Electronic and magnetic properties of dopant atoms in  SnSe monolayer: a first-principles study}
\author{Qingxia Wang}
\affiliation{International Laboratory for Quantum Functional Materials of Henan, and School of Physics and Engineering, Zhengzhou University, Zhengzhou, 450001, China}
\author{Weiyang Yu}
\affiliation{International Laboratory for Quantum Functional Materials of Henan, and School of Physics and Engineering, Zhengzhou University, Zhengzhou, 450001, China}
\affiliation{School of Physics and Chemistry, Henan Polytechnic University, Jiaozuo, 454000, China}
\author{Xiaonan Fu}
\affiliation{Department of Physics and School of Science, Henan University of Technology, Zhengzhou 450001, China}
\author{Chong Qiao}
\affiliation{International Laboratory for Quantum Functional Materials of Henan, and School of Physics and Engineering, Zhengzhou University, Zhengzhou, 450001, China}
\author{Congxin Xia}
\email[e-mail address:]{xiacongxin@htu.edu.cn}
\affiliation{Department of Physics,Henan Normal University, Xinxiang, 453000, China}
\author{Yu Jia}
\email[e-mail address:]{jiayu@zzu.edu.cn}
\affiliation{International Laboratory for Quantum Functional Materials of Henan, and School of Physics and Engineering, Zhengzhou University, Zhengzhou, 450001, China}
\date{\today}
\begin{abstract}

\textbf{Abstract:}
  SnSe monolayer with orthorhombic \emph{Pnma} GeS structure is an important two-dimensional (2D) indirect band gap material at room temperature. Based on first-principles density functional theory calculations, we present systematic studies on the electronic and magnetic properties of X (X = Ga, In, As, Sb) atoms doped SnSe monolayer. The calculated electronic structures show that Ga-doped system maintains semiconducting property while In-doped SnSe monolayer is half-metal. The As- and Sb- doped SnSe systems present the characteristics of $n$-type semiconductor. Moreover, all considered substitutional doping cases induce magnetic ground states with the magnetic moment of ~1$\mu_B$. In addition, the calculated formation energies also show that four types of doped systems are thermodynamic stable. These results provide a new route for the potential applications of doped SnSe monolayer in 2D photoelectronic and magnetic semiconductor devices.

\textbf{Keywords:} SnSe monolayer, substitutional doping, electronic properties, magnetism

\end{abstract}

\pacs{73.20.At, 75.50.Pp, 75.75.+a}
\maketitle

\textbf{1. Introduction}

Along with the discovery of graphene, many new two-dimensional(2D) atomic layered systems have attracted special attention for future electronics applications.\cite{Novoselov,Geim,Rodin,Brumme,Sivek} Recently, more and more researchers devote themselves on 2D atomic-layer materials, such as silicene, germanane, stanene, phosphorene, and so on.\cite{Yu,Zhu1,Vogt,Xia,Zhu2} Just as the scope of group IV semiconductors such as graphene and silicene have been broadened significantly by introducing isoelectronic III-V compounds, phosphorene has been broadened by introducing isoelectronic IV-VI compounds.\cite{Zhu3} IV-VI group semiconductors have layered structures which are similar to phosphorene.\cite{Rarenteau, Choi} For bulk Tin selenium (SnSe), it has two most common crystal structures: one is cubic NaCl (rock salt) and the other is orthorhombic GeS structure with an orthorhombic \emph{Pnma} space group at room temperature.\cite{Lefebvre,Chamberlain} Furthermore, bulk SnSe is an important double layered binary IV-VI semiconductor with structural orthorhombic symmetry and weak van der Waals force between double layers. In addition, bulk SnSe has both a direct band gap of 1.30 eV and an indirect band gap of 0.90 eV, just falling into the optimum band gap for solar cells. Therefore, bulk SnSe is also considered as the most promising candidates for solar cells.\cite{Antunez,Liu,Baumgardner} Especially, with the successfully synthesizing of 2D SnSe experimentally, SnSe monolayer has attracted more and more attention in monolayer IV-VI semiconductor.\cite{Li1} SnSe monolayer is comprised of zigzag double-layered planes and exhibits strong anisotropic properties.\cite{Antunez,Pejova,Xue1} As material of earth-abundance, environment friendly and good chemical stability, SnSe can be applied in a wide range of potential applications, e.g. optical devices, memory switching devices, infrared optoelectronic devices and anode materials for rechargeable lithium batteries.\cite{Chun,Zhang1,Xiao,Xue2,Zhang2}

SnSe monolayer has the similar structure to phosphorene. Sn and Se atoms present a puckered surface due to the $sp^2$ hybridization.\cite{Li2} Each Sn atom forms three covalent bonds with Se, and the same as Se atom. The synthesis and properties of SnSe monolayer have attracted much attention.\cite{Antunez,Li1} As we know, 2D material have many special properties due to their unique dimension-dependent properties comparing with 3D material.\cite{Rogers} Whereas there is few literature about 2D SnSe monolayer,\cite{Zhu3} the electronic and magnetic properties of substitutional doping of SnSe monolayer have not been done. In this work, using first-principles density functional theory (DFT) calculations, we present a systematic study of the electronic and magnetic properties of Ga-, In-, As-, and Sb- doped SnSe monolayer. Firstly, we calculate the electronic and magnetic properties of the doped systems. And then, we calculate the formation energy $E_{f}$ and find that the formation energy of Ga-doped system is lowest than the other three systems,indicating that Ga-doped system is apt to be realized in experiment.

\textbf{2. Computational Details}

The present first-principles DFT \cite{Kohn} calculations were performed using Perdew-Burke-Ernzerh (PBE)\cite{Perdew} in the projector augmented wave (PAW) \cite{Blochl} within Vienna \emph{ab initio} Simulation Package (VASP) .\cite{Kresse} The reciprocal space was sampled with a fine grid of 3$\times$3$\times$3 $k$-points in the Brillouin zone of the primitive unit cell for bulk SnSe, and 1$\times$3$\times$3 $k$-points for the SnSe monolayer and the doped systems. The wave functions were expanded in a plane-wave basis with an energy cutoff of 500 eV. The maximum force of each atom was less than 0.01 eV/{\AA}.The geometrical structure was employed with a 3$\times$3 supercell including 36 atoms with doping concentration of 2.78\%. A vacuum spacing perpendicular to the plane was employed to be at least 15{\AA} in the unit cell to avoid the coupling between neighboring cells. The spin-polarized was considered to analyze the magnetism of the doped systems.

\textbf{3. Results and discussion}

Fig. 1 presents the geometrical structure and electronic band structure of bulk SnSe. The calculated optimal lattice parameters of bulk SnSe are a=11.50{\AA} b=4.16{\AA}, and c=4.45{\AA}, which are in agreement with the experiment values\cite{Franzman} and previous theoretical values\cite{Kutorasinski,Huang}. The bond lengths of Sn-Se are  $d_1$=2.81{\AA} and  $d_2$=2.78{\AA}, which are coincided with the previous reported values,\cite{Lefebvre} as seen in Fig. 1 (a). The calculated band gap is 0.50 eV, which is smaller than experimental value and the previous theoretical calculation,\cite{Huang} on account of the DFT-PBE method which underestimates the band gap of semiconductor. Even though the fundamental band gaps are typically underestimated in DFT-PBE approach, the prediction of the indirect semiconductor system is correct.\cite{Lu,Krasheninnikov,Ramasubramaniam,Cheng} In Fig. 1 (b), we can see that the valence band maximum (VBM) of bulk SnSe is at M point, while conduction band minimum (CBM) is located at the middle of $\Gamma$-M, indicating an indirect band gap. Moreover, the bulk SnSe has no magnetism according to our calculation.

The structure of SnSe monolayer is displayed in Fig. 2, along with the similar structure of phosphorene. As seen in Fig. 2 (a) and (b), Sn and Se atoms form three covalent bonds with each other in a puckered surface, which is similar to that of phosphorene,\cite{Yu,Zhu1} as shown in Fig. 2 (c) and (d). However, it is obvious to perceive that the structure of phosphorene is of higher symmetry than SnSe monolayer, so they appear some different properties, e.g. SnSe monolayer is an indirect band gap semiconductor, while phosphorene has a direct band gap.\cite{Yu}

\begin{figure}[t]
\includegraphics[width=7cm]{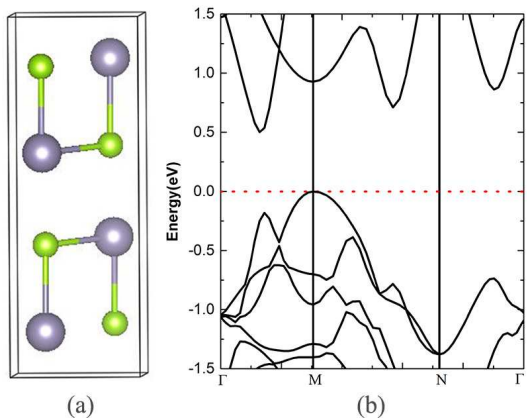}
\caption{(a) The geometric and (b) electronic structures of bulk SnSe.}
\end{figure}

To investigate the electronic and magnetic properties of substitutional doping in  SnSe monolayer, the band structure, total density of states (DOS) and partial density of states (PDOS) of the doping systems $Sn_{17}$$Se_{18}$$X_{Sn}$ (X=Ga, In, As, Sb) are calculated, along with the pure SnSe monolayer for comparison.

\begin{figure}[t]
\includegraphics[width=8cm]{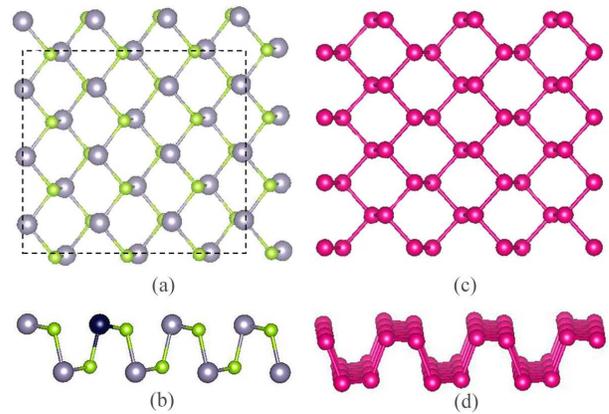}
\caption{(a) Top and (b) side views of SnSe monolayer with 3$\times$3 supercell cell, indicating the black dashed rectangle. The dark atom denotes Sn atom substituted by a dopant atom, and navy balls denote Sn atoms and prasinous balls denote Se atoms. (c) Top and (d) side views of black phosphorene (pink balls denote P atoms).}
\end{figure}

Firstly, band structures, DOS and PDOS of SnSe monolayer are calculated in Fig. 3. As shown in Fig. 3 (a), the indirect band gap is 0.78 eV, which is bigger than that of bulk SnSe ($E_g$=0.50 eV). From the PDOS of SnSe monolayer , as shown in Fig. 3 (b), we can clearly see that it is ${5s}$-orbitals of Sn and ${4p}$-orbitals of Se that contribute mostly to the total DOS below Fermi level from -1.50 eV to -0.11 eV. While above Fermi level, ${5p}$-orbitals of Sn and ${4s}$-orbitals of Se contribute mostly to the total DOS. In detail, it is the hybridization of ${5s}$-orbital of Sn and ${4p}$-orbital of Se that forms the valence band, and ${5p}$-orbital of Sn and ${4s}$-orbital of Se hybridize to comprise the conduction band. The spin-polarized calculations show that pristine SnSe monolayer is of non-magnetism.

\begin{figure}[t]
\includegraphics[width=8cm]{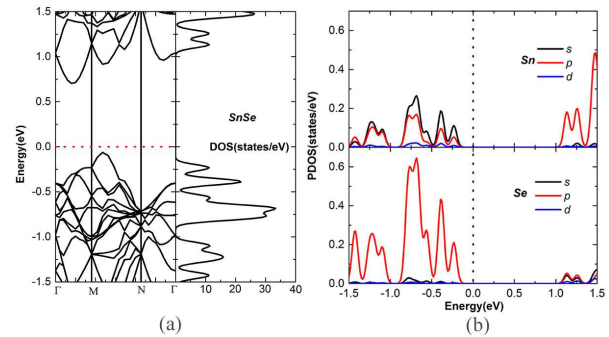}
\caption{(a) Electronic band structure along $\Gamma$-M-N-$\Gamma$ direction and corresponding total DOS, (b) PDOS of SnSe monolayer. The energy zero represents the Fermi level. }
\end{figure}

\begin{figure*}[t]
\includegraphics[width=15cm]{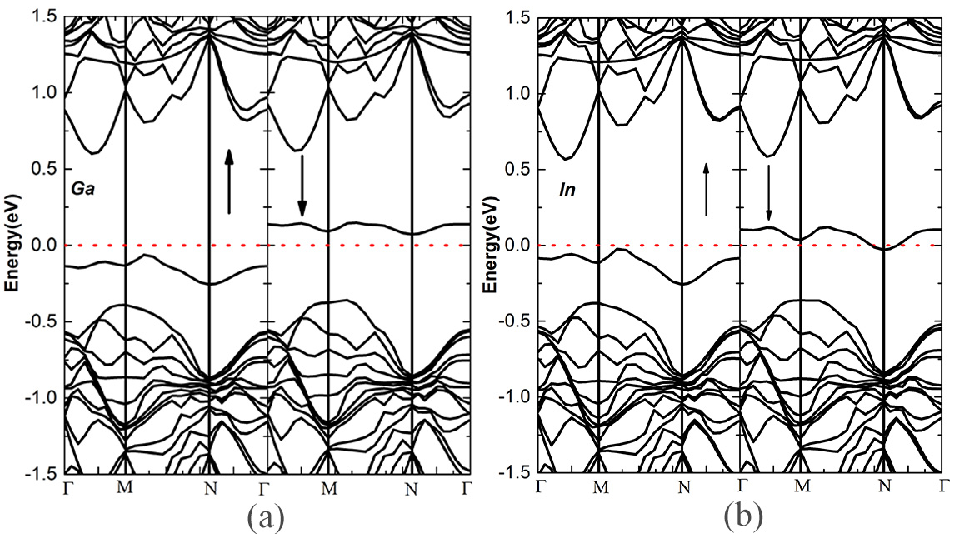}
\caption{Calculated band structures of (a) Ga- and (b) In-doped systems, along $\Gamma$-M-N-$\Gamma$ direction. The energy zero represents the Fermi level.}
\end{figure*}

\begin{figure*}[t]
\includegraphics[width=15cm]{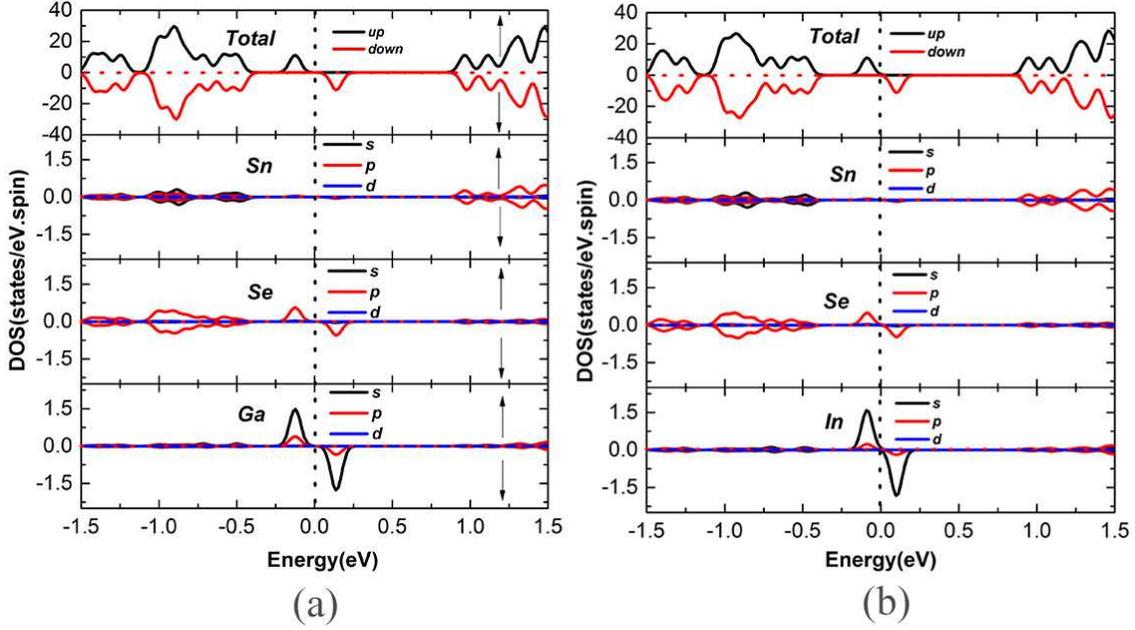}
\caption{DOS and PDOS of (a) Ga-doped and (b) In-doped systems. The energy zero represents the Fermi level. }
\end{figure*}

Now we turn to investigate the doping effects on electronic structures of SnSe monolayer. As we know that the electronic structure of SnSe monolayer can be tuned in a widely range by substitutional doping with Sn atom replaced by other atoms because the outer electron valence configuration of Sn is $4d^{10}$$5s^{2}$$5p^{2}$. There are two $5p^{2}$ electrons of Sn which join in bonding, and two $5s^{2}$ electrons left to form a lone pair. As a result, Sn atom can form s$p^{2}$ bonding with a lone pair of valence electrons.\cite
{Chamberlain} In the doped system $Sn_{17}$$Se_{18}$$X_{Sn}$ (X=Ga, In, As, Sb), X dopant atom is neighborhood with Sn in the periodic table of elements, which can avoid of lattice distortion, giving rise to the impurity states always appear either above valence band or below conduction band, or the middle of band gap. To explore the underlying electronic and magnetic properties of the doped monolayer $Sn_{17}$$Se_{18}$$X_{Sn}$ (X=Ga, In, As, Sb). We divide the related figures into two parts: spin up and spin down, which makes us to have a clear insight to the substitutional doping systems.

Fig. 4 (a) presents the split band structure of Ga-doped system. The impurity state is split into spin-up and spin-down cases, and the impurity energy level appears either above valence band (spin-up) or below conduction bands (spin-down). Furthermore, the band gap is 0.66 eV for spin-up and 0.43 eV for spin-down, indicating semiconductor properties. The \emph{Eg} values of spin-up and spin-down are both smaller than that of pure SnSe monolayer. Since Ga is the acceptor, leading to the band gap decreasing both for spin-up and spin-down. When Ga atom is doped, the doped system has a red shift, giving rise to a relatively wide application in electron devices. However, for In-dpoed system, it is exceptive that the VBM rises for spin-up, while the CBM goes down cross the Fermi level for spin-down, which lead to a half-metal property, as shown in Fig. 4 (b). For the Ga- and In-doped systems, because Ga and In belong to the same group IIIA, they present different properties. The intriguing phenomenon can be explained by the nearly equal electronegativity (Ga$\sim$1.81, Sn$\sim$1.96, In$\sim$1.78), and we know that the difference between Sn and In is slightly bigger than that between Sn and Ga. To have a further insight into the underlying mechanism of the dopant systems, the total DOS and PDOS of Ga- and In-doped system are shown in Fig. 5.

\begin{figure}[t]
\includegraphics[width=8cm]{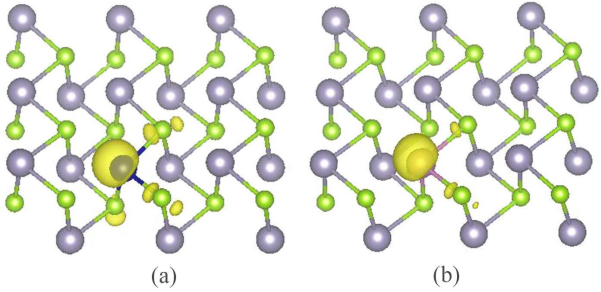}
\caption{Spin charge density ($\rho$$_{up}$-$\rho$$_{down}$) of (a) Ga- and (b) In-doped system (dark blue ball denotes Ga atom, light pink ball denotes In atom),respectively. The value of isosurfaces is 0.03 e/${\AA^3}$. }
\end{figure}

From the PDOS of Ga-doped system, as seen in Fig. 5 (a), we can see that it is obvious that 4$s$-, 4$p$-orbital of Ga and 4$p$-orbital of Se play an important role for the VBM and CBM, and make the main contribution to the total DOS. Similarly, the In-doped system presents the almost same phenomenon: 5$s$-, 5$p$-orbital of In and 4$p$-orbital of Se play an important role for VBM, CBM, and the total DOS. Comparing to the semiconductor property of Ga-doped system, it is obvious that the 5$p$-orbits of In atom presents hybridization at the Fermi level. Consequently, the doping system presents a half-metallic property, which is consistent with the band structure. Moreover, we know that the total DOS is up and down asymmetry nearby the Fermi level, indicating that Ga- and In-doped system present magnetism with the values about 1.00 $\mu_B$ and 0.99 $\mu_B$, respectively, which are close to their free atoms (Ga$\sim$1.00 $\mu_B$; In$\sim$1.00 $\mu_B$). Furthermore, from Fig. 5 (a), it is obvious that the majority of magnetism is induced by Ga atom while Se atom contributes minority to the magnetism. But in In-doped system, the magnetism is mainly induced by the In dopant, as seen in Fig. 5 (b).

In order to get a clear explanation of magnetism, the spin charge density of Ga- and In- doped systems are calculated in Fig. 6. From Fig. 6 (a) and (b) we can obviously see that the magnetism is induced by the dopant Ga atom and In atom, respectively. The phenomena are coincide well with the results of DOS and PDOS. And in the Ga-doped system, the majority of magnetism distributes around the Ga atom while the minority scatters around Se atom. Comparing to Ga-doped system, the magnetism is mainly induced by the dopant atom in In-doped system. The reason of the magnetism is that when Ga and In replaced Sn in SnSe monolayer, the system will exist a hole because of the formed covalent band. Hence, the hole will induce the magnetism. As a result, both Ga and In dopant atoms lead to an acceptor level which gives rise to $p$-type doping systems with band gap decreasing.

\begin{figure*}[t]
\includegraphics[width=15cm]{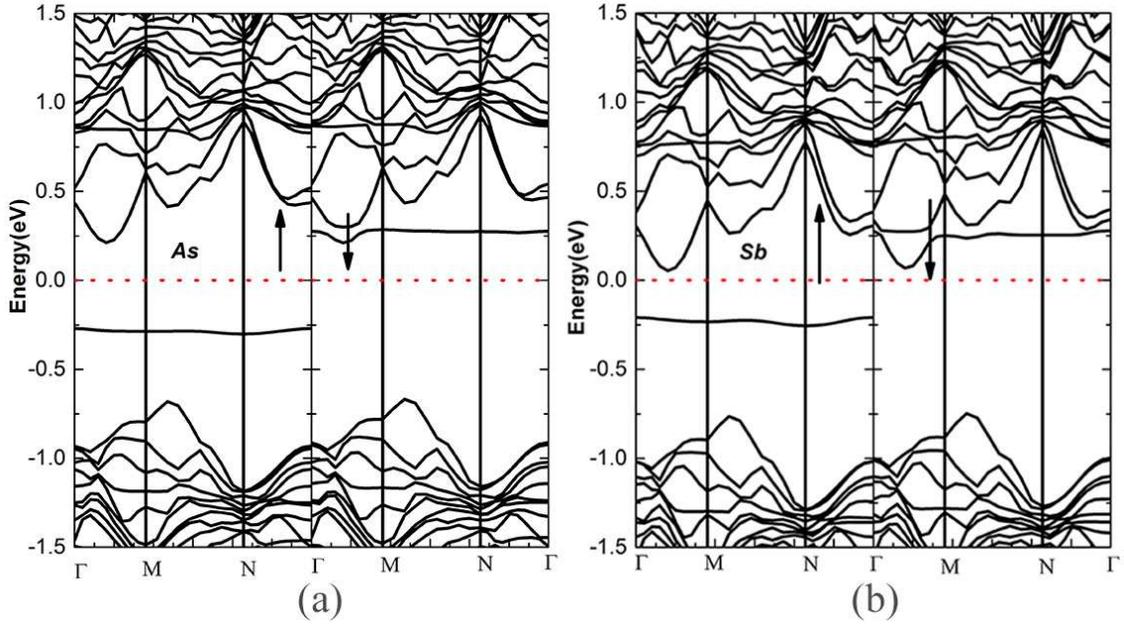}
\caption{Calculated band structures of (a) As- and (b) Sb-doped systems, along $\Gamma$-M-N-$\Gamma$ direction. The energy zero represents the Fermi level.}
\end{figure*}

\begin{figure*}[t]
\includegraphics[width=15cm]{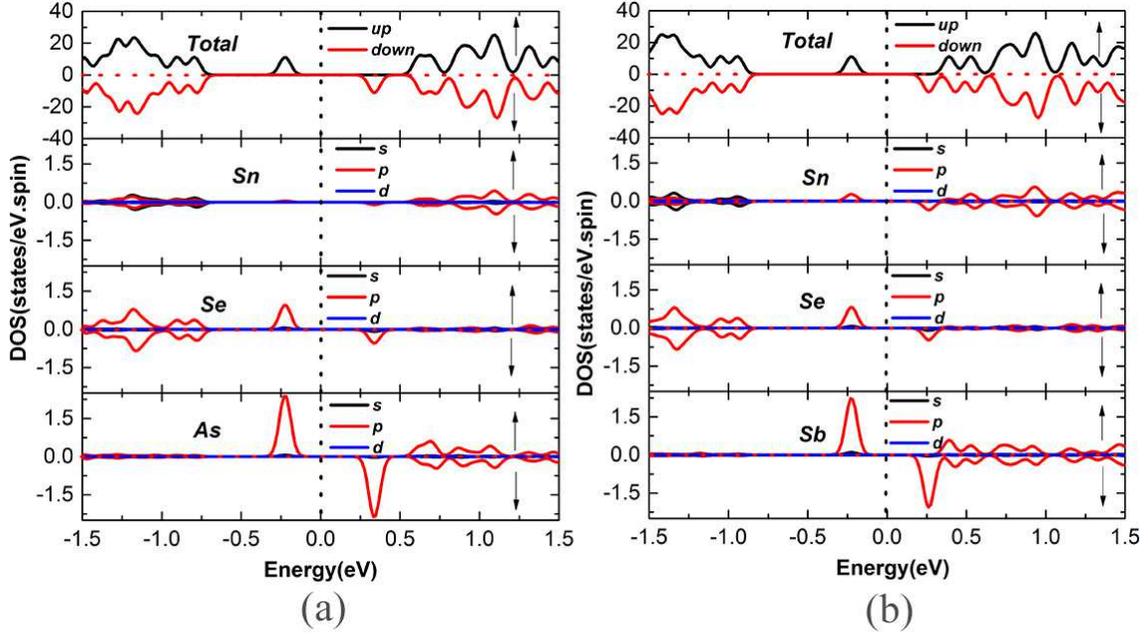}
\caption{DOS and PDOS of (a) As- and (b) Sb-doped systems. The energy zero represents the Fermi level.}
\end{figure*}

\begin{figure}[t]
\includegraphics[width=8cm]{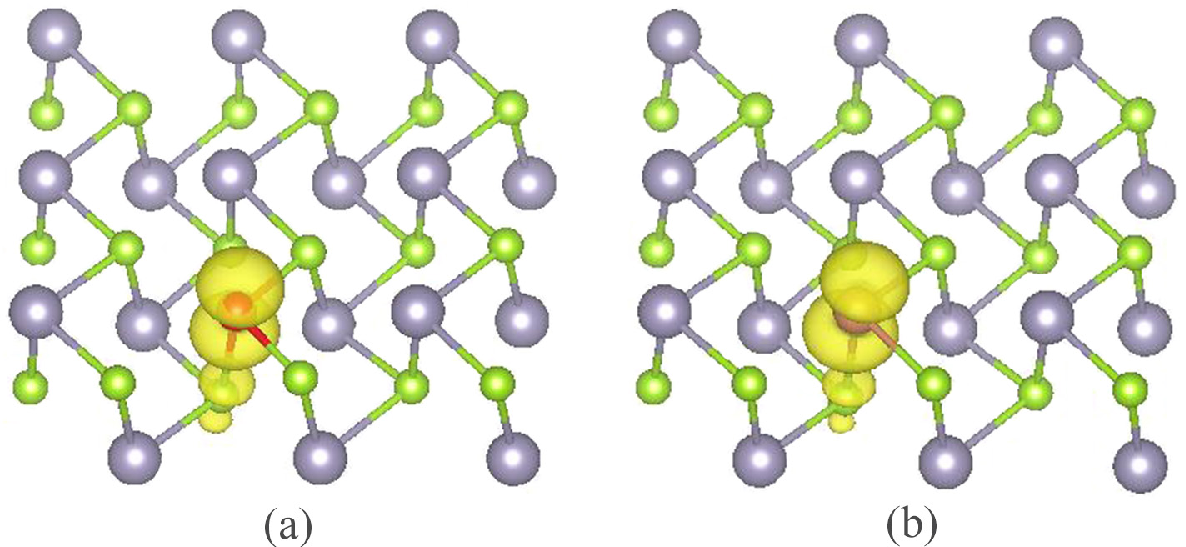}
\caption{Spin charge density($\rho$$_{up}$-$\rho$$_{down}$) of (a) As- and (b) Sb-doped system  (red ball denotes As atom, brown ball denotes Sb atom), respectively. The value of isosurface is 0.03 e/${\AA}^3$.}
\end{figure}

In addition, we calculate the electronic band structure of As- and Sb-doped systems in Fig. 7. From Fig. 7 we can see that the impurity level appears in valence bands leading to the VBM rise for spin-up, while in conduction bands resulting in the CBM declining for spin-down. However, the impurity level does not cross the Fermi level, so the system maintains the semiconductor property for As- and Sb-doped systems. In detail, for the As-doped system, as seen in Fig. 7 (a), the band gap is divided into two parts: 0.48 eV for spin-up and 0.88 eV for spin-down. At the same time the band gap of Sb-doped system is 0.26 eV for spin-up and 0.84 eV for spin-down, as seen in Fig. 7 (b). Otherwise, comparing to the band gap of SnSe monolayer, we find that the value of Sb-doped system has a less change than that of As-doped system. This phenomenon can be explained by electronegativity (As$\sim$2.18; Sn$\sim$1.96; Sb$\sim$2.05). It is visible that the electronegativity difference between Sn and As is larger than that between Sn and Sb. In order to have a further penetration into the electronic property of dopant systems, the total DOS and PDOS of As- and Sb-doped are shown in Fig. 8.

From Fig. 8 (a) we can obviously see that 4$p$-orbitals of As makes the contribution to not only VBM and CBM, but also the total DOS. In other words, the main contribution is from 4$p$-orbitals of the dopant As, which is coincide with the explanation by band structure. Furthermore, there is a flat-level for the band structure in spin-up, leading to a large density of states in that region, as shown in PDOS of As. In addition, we find that the system is not symmetry at the Fermi level in total DOS, which exhibits the system is of magnetism. Obviously, the magnetism is mainly induced by As atom. Then we calculated the magnetic value (1.00 $\mu_B$), which is smaller than its free atom (As$\sim$3.00 $\mu_B$). On the other hand, for Sb-doped system, as seen in Fig. 8 (b), it shows up the similar properties to As-doped system that 4$p$-orbitals of Sb play an important role for VBM, CBM and total DOS. The total DOS appears asymmetry at the Fermi level, indicating the Sb-doped system is of magnetism. Obviously, the magnetism is mainly derived from the Sb atom, and the value of calculated magnetism is 1.00$\mu_B$, which is also smaller than its free atom (Sb$\sim$3.00$\mu_B$). From Fig. 8, we obtain the same explanation with that of the band structures. Finally, in order to explain the magnetic mechanism, we calculated the spin charge density of As- and Sb-doped systems as shown in Fig. 9.

From Fig. 9 (a) and (b), we can distinctly see that the magnetism is mainly induced by As atom and Sb atom for As- and Sb-doped systems, respectively. For the As-doped system, the magnetism mainly gathers around the As atom, and minority scatters around Se atoms. Similarly, the magnetism is mainly induced by the dopant atom in Sb-doped system. The phenomena are in line with the results of DOS and PDOS. The reason of the magnetism is that when As and Sb replaced Sn in SnSe monolayer, the system will exist an extra electron. Hence, the extra electron will induce the magnetism. As a result, both As and Sb dopant atoms lead to an donor level resulting in $n$-type doped systems.

Generally, the X-doped (X=Ga, In, As, Sb) systems with either half-metal or semiconductor properties can be explained by the $sp^{2}$ bonding character of Sn atoms with a lone pair of valence electrons and their strong hybridizations with the \emph{sp} orbitals of dopants. We can conclude that the doped systems $Sn_{17}$$Se_{18}$$X_{Sn}$ (X=Ga, In, As, Sb) are ferromagnetic states from the spin charge density. Considering our work, we are aware that the electronic properties can be tuned by substitutional doping.

\begin{table}
\caption{Calculated structural and magnetic properties for single atoms doping on SnSe monolayer. Formation energy ($E_{f}$); the magnetic moments($\mu$); minimum dopants-Se (D-Se) distance($d_{2}$).}
\begin{tabular}{p{2cm}p{2.0cm}p{1.5cm}p{2.0cm}}
\hline
\hline
Dopant         & $E_{f}$(eV/atom)   &$\mu$ ($\mu_{B}$)    &$d_{2}$ (D-Se) (\AA)\\
\hline
SnSe           &  -4.13             &  0                  &  2.81 \\
Ga             &  -2.57             &  1.00               &  2.55 \\
In             &  -2.54             &  0.99               &  2.79 \\
As             &  -1.87             &  1.00               &  3.08 \\
Sb             &  -1.91             &  1.00               &  3.17 \\
\hline
\hline
\end{tabular}
\end{table}

Finally, we calculate the formation energy of the substitutional system in order to verify the stability of X-doped SnSe monolayer. The formation energy ($E_{f}$) is defined as $E_{f}$ = $E_{dh}$ - $E_{d}$ - $E_{h}$, where $E_{dh}$, $E_{d}$, and $E_{h}$ are the total energies of the doped system $Sn_{17}$ $Se_{18}$ $X_{Sn}$ (X=Ga, In, As, Sb), the energy of monolayer $Sn_{17}$ $Se_{18}$ and single Sn atom and isolated X atoms, respectively. From our definition, the negative value of $E_{f}$ shows that a system is stable. The large absolute value of $E_{f}$ energy means strong interaction between dopants and Sn. A summary of the results is shown in Table 1. From Table 1, we find that the formation energy of Ga-doped system is minimum, which demonstrates Ga atom is easy to be doped and the Ga-doped system is the most stable than other three systems. In general, the doped systems can be realized in experiment.

\textbf{4. Conclusions }

In summary, using first-principles calculations, we present the geometrical structure, electronic and magnetic properties of monolayer SnSe substitutionally doped by Ga, In, As, and Sb atoms, respectively. We find that the electron structure can be tuned by substitutional doping. When Ga or In is doped, the system presents semiconductor and half-metal properties, respectively; while As or Sb substitute Sn, the systems display semiconductor properties. In Ga-doped system, due to the band gap declining, the system exhibits red shift, which makes the system to have a wide application in optoelectronic devises. When Sn atom is replaced by Ga atom or In atom, inducing a hole, so the systems belong to $p$-type doping; on the other hand, if As or Sb replace Sn, the system has an unpaired electron, resulting $n$-type doping. No matter whether Ga and In as dopant atoms, or As and Sb, there is an unpaired electron which can induce about 1$\mu_B$ magnetism. Finally, comparing to the formation energy of $Sn_{17}$$Se_{18}$$X_{Sn}$ (X=Ga, In, As, Sb), we find that Ga is easy to be doped and the Ga-doped system of is the most stable, whereas the four types of doped systems are thermodynamic stable. The results of substitutional doping make the application of SnSe monolayer more extensive.

\textbf{5. Acknowledgements}

This work is supported by financial support from the National Basic Research Program of China (No.2012CB921300) and the National Natural Science Foundation of China (Grant Nos.11274280).

\end{document}